\documentclass[twocolappendix]{emulateapj}

\usepackage{soul,color,amsmath,comment}

\shorttitle{Rayleigh-Taylor Turbulent GRB Afterglows}
\shortauthors{Duffell \& MacFadyen}

\begin{document}

\title{Shock Corrugation by Rayleigh-Taylor Instability in GRB Afterglow Jets}

\author{Paul C. Duffell and Andrew I. MacFadyen}
\affil{Center for Cosmology and Particle Physics, New York University}
\email{pcd233@nyu.edu, macfadyen@nyu.edu}

\begin{abstract}

Afterglow jets are Rayleigh-Taylor unstable and therefore turbulent during the early part of their deceleration.  There are also several processes which actively cool the jet.  In this letter, we demonstrate that if cooling significantly increases the compressibility of the flow, the turbulence collides with the forward shock, destabilizing and corrugating it.  In this case, the forward shock is turbulent enough to produce the magnetic fields responsible for synchrotron emission via small scale turbulent dynamo.  We calculate light curves assuming the magnetic field is in energy equipartition with the turbulent kinetic energy and discover that dynamic magnetic fields are well-approximated by a constant magnetic-to-thermal energy ratio of $1\%$, though there is a sizeable delay in the time of peak flux as the magnetic field turns on only after the turbulence has activated.  The reverse shock is found to be significantly more magnetized than the forward shock, with a magnetic-to-thermal energy ratio of order 10\%.  This work motivates future Rayleigh-Taylor calculations using more physical cooling models.

\end{abstract}


\section{Introduction}
\label{sec:intro}

Magnetized relativistic jets are important astrophysical phenomena, most notably in the context of gamma ray bursts (GRBs), but also in active galactic nuclei and tidal disruption events.  As a result, the dynamics of relativistic jets have been studied extensively, often in terms of the GRB central engine \citep{1999ApJ...524..262M, 2000ApJ...531L.119A, 2007ApJ...665..569M, 2009MNRAS.397.1153K, 2012MNRAS.423.3083M, 2013ApJ...767...19L}, but also in the largely engine-independent afterglow phase when ejecta accelerated by the central engine has transferred its energy to a collimated blast wave \citep{1999ApJ...525..737R, 2000ApJ...541L...9K, 2001grba.conf..312G, 2002ApJ...571..779P, 2002bjgr.conf..146L, 2003ApJ...586..356Z, 2005ApJ...626..966P, 2007RMxAC..27..140G, 2010ApJ...722..235V, 2012ApJ...751...57D}.  Given these extensive studies, there are still many fundamental questions which remain unanswered.  For example, afterglow jets are thought to be magnetized, as synchrotron emission necessitates a strong magnetic field, yet no clear mechanism has been demonstrated which robustly generates such a field.  Additionally, current jet models are parameterized by a small handful of parameters \citep{2012ApJ...747L..30V}, which would seem to suggest a straightforward standardization of GRB afterglow light curves.  However, GRB afterglows display a great deal of variety and variability, especially at early times, hence there likely exist additional important elements missing from simplified hydrodynamical models.

One avenue which potentially addresses these issues is vorticity generation behind the forward shock.  Vorticity could both amplify magnetic fields via turbulent dynamo and produce variability in GRB light curves.  Understanding where vorticity comes from and how much is present will help to complete the picture of how relativistic jets generate afterglow emission.

The source of vorticity is still unclear, but several mechanisms have been suggested.  One possibility is small-scale Weibel instabilities in the plasma particles making up the shock itself \citep{2008ApJ...673L..39S}.  However, such instabilities may have a short range of influence.  Alternatively, vorticity can be generated when a shock overtakes high-density clumps in the interstellar medium (ISM) \citep{2007ApJ...671.1858S, 2008JFM...604..325G}, but it is unclear whether large enough clumps exist to make this a robust mechanism.

In this work, we consider the vorticity generated by Rayleigh-Taylor (RT) instability, as first suggested by \cite{2009ApJ...705L.213L}.  After a GRB ejects a relativistic flow (ejecta), it expands and its thermal energy drops adiabatically until it is subdominant to the kinetic energy.  The ejecta then coasts and becomes a very thin shell with width $\Delta r / r \sim 1 / \Gamma^2$, where $\Gamma$ is the Lorentz factor \citep{1999ApJ...513..669K}.  When deceleration finally occurs, shocks are generated at the interface between ejecta and ISM.  A forward shock pushes its way into the ISM, and a reverse shock pushes its way back into the ejecta.  In the heated region between these two shocks resides the contact discontinuity, separating ejecta from ISM.  This contact discontinuity is Rayleigh-Taylor unstable.


Nonrelativistic RT-unstable outflows were first studied by \cite{1992ApJ...392..118C}, both analytically and numerically.  \cite{1996ApJ...465..800J} later performed a two-dimensional magnetohydrodynamics calculation which demonstrated how magnetic fields tend to align themselves along RT fingers.  More recently, \cite{2010AnA...509L..10F} and \cite{2011MNRAS.415...83W} have demonstrated the importance of various microphysical processes at the shock front, and \cite{2010AnA...515A.104F} has performed 3D numerical calculations.  To extend the nonrelativistic results into the relativistic regime, \cite{2010GApFD.104...85L} performed a stability analysis on the two-shock solution \citep{2006ApJ...645..431N} and found linear growth rates which could potentially be large enough to impact the forward shock.

In the first numerical studies of the relativistic case, \cite{2013ApJ...775...87D} found that Rayleigh-Taylor generates turbulence which could amplify magnetic fields to within a few percent of equipartition with the thermal energy density.  However, in that work we found the turbulence remained confined within a region behind the forward shock and did not impact the forward shock, though turbulence did penetrate part of the energetic post-shock region.
\\

In this letter we demonstrate that it is possible for the Rayleigh-Taylor turbulence to collide with the forward shock.  As a result, the shock is perturbed and corrugated and significant turbulence is present everywhere behind it.  This turbulence persists for a long time, until the shock becomes nonrelativistic, possibly due to the non-universality of the Blandford-McKee solution \citep{2000astro.ph.12364G}.

The key ingredient allowing the turbulence to collide with the forward shock is a softer equation of state.  In fact, a softened equation of state has already been invoked in the nonrelativistic case to explain how Rayleigh-Taylor fingers can catch up to the forward shock in Type 1A supernovae \citep{2001ApJ...560..244B, 2010AnA...509L..10F, 2011MNRAS.415...83W}.  In this case, the collision of the ejecta with the forward shock is apparent in images of supernova remnants \citep{2005AnA...433..229V, 2011AnA...532A.114K, 2012SSRv..173..369H}.

A softened equation of state can result in a reduced pressure gradient in the forward shock.  This pressure gradient acts as a restoring force keeping the Rayleigh-Taylor fingers behind the forward shock, so if the pressure gradient is reduced significantly, the turbulence can collide with the forward shock.  Therefore, if cooling removes a non-negligible fraction of the internal energy, it can reduce this pressure gradient, facilitating the collision of the turbulence with the shock.

There are several reasons that the equation of state of GRB jets is expected to be softer than an adiabatic $4/3$ law.  Cosmic ray acceleration at the forward shock can carry a significant amount of thermal energy, cooling the shock \citep{2011AnA...532A.114K, 2012ApJ...749..156O}, which may effectively result in a softer equation of state.  Additionally, the shock is highly radiative, so that photon production also provides cooling.

Photon cooling can potentially impact the dynamics; for example, GRB 080319B was estimated to emit $\sim 10^{51}$ ergs in X-rays \citep{2008Natur.455..183R, 2009ApJ...691..723B}, which should be a non-negligible fraction of the energy in the blastwave.  If this cooling is responsible for reduced pressure in the shock front, this might require some coupling between the leptons and the baryons.  The cooling from cosmic rays has less certain observational constraints in the GRB context, but it has been found to be important dynamically for the nonrelativistic case of supernova remnants \citep{2011AnA...532A.114K, 2012APh....39...33A}.

In order to elucidate the effects of increased compressibility, we compare two Rayleigh-Taylor setups differing only in the adiabatic index: the usual relativistic $\gamma = 4/3$, and $\gamma = 1.1$ representing the case where cooling is dynamically important.  It is straightforward to see that this reduced index should result in a lower pressure for fixed internal energy, since $P = (\gamma-1)\epsilon$.  The difference between $4/3$ and $1.1$ can be envisioned as the difference between $P = \epsilon/3$ and $P = \epsilon/10$, so that for a given internal energy the pressure is reduced by about a factor of $3$.  Such a change can be roughly interpreted as losing $2/3$ of the thermal energy to cooling.  Therefore, this adiabatic index models a system which loses a non-negligible fraction of its internal energy.  As we shall see, the reduced pressure allows Rayleigh-Taylor fingers to impact the forward shock, and generate plenty of vorticity for the entire time the shock is relativistic.  Thus, the shock will continue to be corrugated as long as the relevant cooling processes are effective at softening the equation of state.  The choice of $\gamma = 1.1$ to represent effects of cooling is a proof-of-concept which motivates further study using a more accurate cooling prescription.
\\

\section{Numerical Set-Up}
\label{sec:numerics}

Our study entails numerically integrating the equations of relativistic hydrodynamics,

\begin{equation}
\partial_{\mu} ( \rho u^{\mu} ) = 0
\end{equation}
\begin{equation}
\partial_{\mu} ( ( \rho + \epsilon + P ) u^{\mu} u^{\nu} + P g^{\mu \nu} ) = 0
\end{equation}
where $\rho$ is proper density, $P$ is pressure, $\epsilon$ is the internal energy density, and $u^{\mu}$ is the four-velocity.  We employ an adiabatic equation of state:
\begin{equation}
P = (\gamma - 1) \epsilon
\end{equation}

and we use relativistic units such that $c = 1$.  

We write the equations in spherical coordinates, and assume axisymmetry so that our calculation is two-dimensional (2D).  Three-dimensional (3D) effects may also be important, as in the nonrelativistic case it has been found that the instability's growth is 30\% larger in 3D than in a 2D calculation \citep{2010AnA...515A.104F}.  Thus, the 3D case will be an interesting complement to this work which we plan to address in the future.

\begin{figure*}
\epsscale{1.0}
\plotone{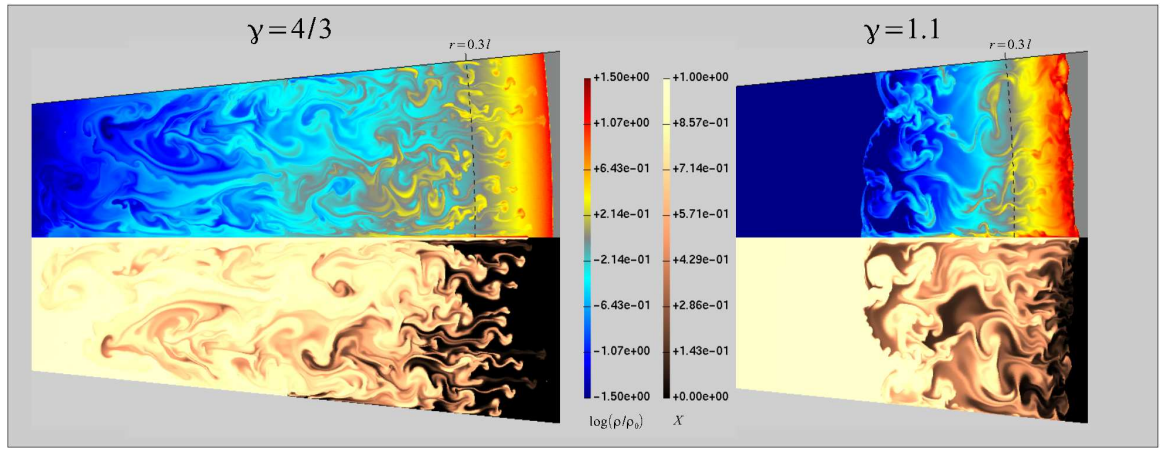}
\caption{ Snapshots of RT turbulence at time $t = l/\Gamma^{1/3} = 0.316 ~l$, using adiabatic index $\gamma = 4/3$ (left), and $\gamma = 1.1$ (right).  As is clearly visible in the figure, the Rayleigh-Taylor turbulence does not collide with the forward shock in the $4/3$ case, but it does in the softer $\gamma = 1.1$ case.
\label{fig:pic} }
\end{figure*}

In order to track the ``ejecta" and ``ISM" components of the flow, we also evolve a passive scalar, $X$, according to
\begin{equation}
\partial_\mu ( \rho X u^\mu ) = 0.
\end{equation}
Initially, we choose $X = 0$ in the ISM and $X = 1$ in the ejecta.  This passive scalar is helpful for visualizing the turbulent mixing of ejecta with the ISM (Figure \ref{fig:pic}).

The calculation is performed using a novel moving-mesh code, JET \citep{2011ApJS..197...15D,2013ApJ...775...87D}.  The JET code uses high-resolution shock-capturing methods, and is effectively Lagrangian in the radial dimension due to the radial motion of grid cells.  In this study we use a resolution of $N_{\theta} = 800$ zones in polar angle, (meaning $\Delta \theta = 1.25 \times 10^{-4}$) and roughly $N_r \sim 8000$ zones radially.  Previously we demonstrated accurate convergence of the JET code for the relativistic RT problem \citep{2013ApJ...775...87D}.


\subsection{Initial Conditions}
\label{sec:ics}

The system is parameterized by an explosion energy $E$, ejecta mass $M$, and ISM density $\rho_0$.  We define a characteristic Lorentz factor $\Gamma = E/M$.  For expediency we choose $\Gamma = 30$, and the constants $E$ and $\rho_0$ simply scale out of the problem, due to the scale invariance of the underlying hydrodynamical field equations.  Our initial conditions are of a cold expanding flow with kinetic energy E and mass M pushing its way into an ISM with density $\rho_0$.  It begins long before a significant amount of the ISM has been swept up, at time $t_0 = 0.01 ~l$, where $l \equiv (E/\rho_0)^{1/3}$ is the Sedov length.  This almost totally specifies the problem, except for the overall shape of the ejecta density profile, which we prescribe as follows based on 1D numerical calculations of relativistic fireballs \citep{2013ApJ...776L...9D}:

\begin{equation}
\rho( r , t_0 ) = \left\{ \begin{array}
				{l@{\quad \quad}l}
				{E \over 2 \pi t_0^3}{1 - R/t_0 \over 1 - r/t_0} & r < R  		\\  
    			\rho_0 & \text{otherwise}  		\\
    			\end{array} \right.
    			\label{eqn:hubble}
\end{equation}

\begin{equation}
\vec v( r , t_0 ) = \left\{ \begin{array}
				{l@{\quad \quad}l}
				\vec r / t_0 & r < R  		\\  
    			0 & \text{otherwise}  		\\
    			\end{array} \right.
    			\label{eqn:hubble}
\end{equation}

\begin{equation}
P( r , t_0 ) \ll \rho( r , t_0 )
\end{equation}

where we have defined 
\begin{equation}
R = t_0 \left(1 - {1 \over 4 \Gamma^2}\right).
\end{equation}


Our domain is axisymmetric, and extends from $\theta = 0$ to $\theta = 0.1$ with a reflecting boundary at $\theta = 0.1$.  This angular size was chosen to represent a patch of a spherical outflow.  During early times in the jet's evolution, while the Lorentz factor is larger than the inverse of the opening angle, causality prevents this choice of opening angle from making any difference in the dynamics.  At late times, it is possible that jet spreading introduces an important dynamic to the turbulence, which we do not attempt to capture here.

\section{Results}
\label{sec:results}	

\begin{figure}
\epsscale{1.0}
\plotone{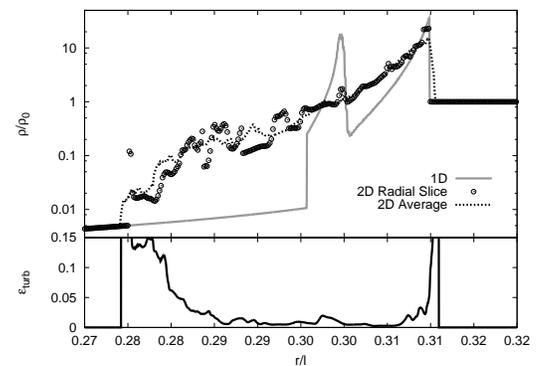}
\caption{ 1D profiles at $t = 0.316 ~l$ for the $\gamma = 1.1$ case.  We plot proper density for a 1D calculation in spherical symmetry, and compare with the 2D version of the calculation.  In 2D, we show the values of proper density along a radial slice at $\theta = 0.05$, as well as a spherically averaged profile.  Additionally, we estimate the magnetic field strength using $\epsilon_{turb}$, the ratio of turbulent energy to thermal energy.
\label{fig:profile} }
\end{figure}

There is a clear difference in the dynamics between the $\gamma = 4/3$ and the $\gamma = 1.1$ case in Figure \ref{fig:pic}.  In the $\gamma = 4/3$ case, the instability collides with the reverse shock, but does not overtake the forward shock.  In the $\gamma = 1.1$ case, the softer equation of state results in lower pressures which allow the Rayleigh-Taylor turbulence to collide with the forward shock, corrugating it and pushing it forward.  The entire heated region between forward and reverse shocks is turbulent.  The corrugated shock front also causes further vorticity generation due to shock obliquity.  In our calculation, we did not see any re-stabilization of the forward shock, which suggests that the turbulence should persist for as long as the soft equation of state is valid.

In Figure \ref{fig:profile} we plot a snapshot of the 1D profile at $t = \Gamma^{-1/3} ~l = 0.316 ~l$.  Here we look at the $\gamma = 1.1$ case, comparing the turbulent 2D calculation with a 1D calculation performed assuming spherical symmetry.  In 1D, we clearly see the forward shock, reverse shock, and contact discontinuity.  In 2D, the contact discontinuity is totally disrupted, and the reverse shock has been pushed back further into the ejecta.  For the 2D results, we plot a spherically averaged proper density, and additionally we plot the density measured along the radial line at $\theta = 0.05$.  We see the turbulent variability exists everywhere between the forward and reverse shocks.  

Turbulence quickly amplifies magnetic fields to rough equipartition with the turbulent kinetic energy density \citep{2003ApJ...597L.141H, 2004ApJ...612..276S,  2012PhRvL.108c5002B, 2013ApJ...769L..29Z}.  Because this turbulence is present all the way up to the forward shock, magnetic fields amplified by the turbulence will facilitate synchrotron emission by the hot electrons in and behind the shock front.

Following the same strategy as in our previous work \citep{2013ApJ...775...87D}, we estimate the magnetic field strength by calculating the energy in turbulent fluctuations.  This is characterized by the assumption
\begin{equation}
\epsilon_B \sim \epsilon_{turb},
\end{equation}
where $\epsilon_B$ is the local ratio of magnetic to thermal energy and $\epsilon_{turb}$ is the local ratio of turbulent to thermal energy.  We calculate this ratio using essentially the same formula as in the previous work:
\begin{equation}
\epsilon_{turb} = { (\gamma - 1)(\langle \rho \rangle_{cons} - \langle \rho \rangle_{vol}) + \langle P \rangle_{cons} - \langle P \rangle_{vol} \over \langle P \rangle_{vol}}
\end{equation}

where brackets denote an average over angle, and the subscript ``vol" implies a simple volume average, whereas ``cons" implies a conservative average (mass, energy, and momentum are averaged, and proper density and pressure are calculated from these conserved quantities).

The turbulent fraction is plotted in the lower panel of Figure \ref{fig:profile}.  We note several important points.  First, the entire region between forward and reverse shocks have $\epsilon_{turb}>0$ and will therefore be magnetized.  Secondly, the smallest values of $\epsilon_{turb}$ are of order 1\%, which by itself is enough to facilitate synchrotron emission.  Third, the largest values of the magnetization are at the forward and reverse shocks, the same place where we hot electrons are expected to produce synchrotron emission.  At the shocks, the magnetization is somewhere between a few percent and ten percent.  Finally, the reverse shock is significantly more magnetized than the forward shock, as $\epsilon_{turb} \sim 0.1$ near the reverse shock, and $\epsilon_{turb} \sim 0.025$ near the forward shock.

\begin{figure}
\epsscale{1.0}
\plotone{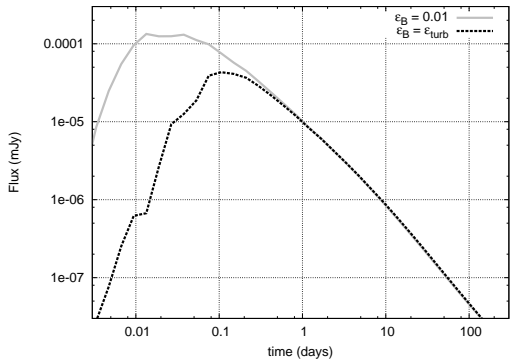}
\caption{ X-Ray afterglow light curves are calculated directly from the output data for adiabatic index $\gamma = 1.1$.  We compare two methods: one assuming a constant magnetic energy fraction $\epsilon_B = 0.01$ and another directly estimating $\epsilon_B$ from the turbulence, equating $\epsilon_B = \epsilon_{turb}$.
\label{fig:afterglow} }
\end{figure}

It is now possible to calculate light curves using an estimate for the magnetic field taken directly from the turbulence calculation, rather than postulating a constant $\epsilon_B$.  To generate an example light curve, we have calculated radiation from the blastwave using a simple synchrotron model.  The synchrotron model is nearly identical to that of \cite{2010ApJ...722..235V}, however we assume an optically thin medium and calculate the flux and observer time for each fluid element, and bin them to calculate a light curve.  Therefore, we do not take into account absorption, but we do model emission, including electron cooling (assuming a global cooling time).  The model requires us to choose specific values for parameters other than $\epsilon_B$, so we choose an isotropic equivalent energy $E_{iso} = 10^{53}$ ergs, an ISM density of 1 proton per $\text{cm}^3$, an electron energy fraction $\epsilon_e = 0.1$, a slope to the electron energy distribution $p = 2.5$, and a luminosity distance of $10^{28}$ cm.  We chose an observer frequency in the X-Ray band, at $10^{18}$ Hz.  First, the 2D profiles are spherically averaged so as to produce a time series of 1D snapshots as in Figure \ref{fig:profile}.  Next, this time-series is fed into our synchrotron model, now assuming the flow is spherically symmetric (we will not see a jet break this way, as that might have made our results somewhat more difficult to interpret).

Sample light curves are plotted in Figure \ref{fig:afterglow}; one assuming a fixed $\epsilon_B = 0.01$, and the other calculated assuming the local formula $\epsilon_B = \epsilon_{turb}$ calculated directly from the averaged fields.  The most remarkable part of Figure \ref{fig:afterglow} is the right-hand side, where the two light curves almost exactly coincide.  This means the magnetic fields generated by Rayleigh-Taylor are very well approximated by a fixed magnetic energy fraction of $\epsilon_B = 0.01$.  We expect, however, that this would not necessarily hold if we used a more realistic cooling model, which stopped impacting the dynamics after some timescale.  In this case, the forward shock could re-stabilize and $\epsilon_{turb}$ would vanish at the shock.  This could also happen during the non-relativistic phase of the afterglow, at which time the adiabatic index would grow to $5/3$, an effect which was not accounted for here.

The second interesting part of this plot is the left-hand side.  In the grey curve corresponding to $\epsilon_B =$ constant, the initial rise is due to the transition from a coasting to decelerating shell.  This peak occurs at time $t_{\gamma} \sim \Gamma^{-2/3} ~l$.  In the case where $\epsilon_B = \epsilon_{turb}$, the dynamics are identical, but the magnetic field does not turn on until a later time, about a factor of $5$ later in observer time, $t_B^{obs} = 5 t_\gamma^{obs}$ (we did not check the scaling with $\Gamma$ as we only performed the $\Gamma = 30$ case).  This means that radiation from an outflow with initial Lorentz factor $\Gamma_0$ will peak at a time $t_{B}$, when the shell has decelerated to a Lorentz factor $\Gamma_B < \Gamma_0$ ($\sim 15$ in our case).  If this peak were interpreted as occurring at $t_{\gamma}$, the Lorentz factor of the ejecta will be incorrectly estimated to be $\Gamma_B$ instead of $\Gamma_0$.  This point may be important for jet models with a baryon-loaded component \citep{1998ApJ...496..311P, 2002MNRAS.337.1349R}, and in general for the interpretation of early-time plateaus in GRB light curves \citep{2006ApJ...642..389N,2006ApJ...642..354Z,2014arXiv1402.5162V}.

\section{Discussion}
\label{sec:disc}

We demonstrate that Rayleigh-Taylor instabilities can generate vorticity and magnetic fields in GRB afterglow jets.  The only ingredient necessary for this mechanism is an equation of state which is softer than the usual $\gamma = 4/3$ model.  This soft equation of state represents a mechanism for energy loss which reduces the pressure in the forward shock so that RT fingers can collide with it.  Several processes occur at the forward shock which act to cool it; cosmic rays and radiation, for example, may carry significant energy away from the shock.  Regardless of what cools the shock front, this seemingly benign change to the dynamics can completely change the structure and magnetization of the blastwave.


We estimate a magnetic energy fraction of $\epsilon_B \sim 1\%$ in the forward shock, and $\sim 10\%$ in the reverse shock.  We show that the choice $\epsilon_B =$ constant $= 0.01$ agrees surprisingly well with late-time afterglow calculated assuming a local value of $\epsilon_B = \epsilon_{turb}$, although this result may change when more accurate models for cooling are employed in the calculation.  Finally, we show that this magnetic field does not turn on until an observer time later than the deceleration time (a factor of $5$ later in our $\Gamma = 30$ case).  This occurs at a time when the shock has decelerated to a Lorentz factor lower than its original value.  This could be related to the cause of observed early-time plateaus in GRB afterglows.

\acknowledgments
This research was supported in part by NASA through Chandra grant TM3-14005X and Fermi grant NNX13AO93G.  

Resources supporting this work were provided by the NASA High-End Computing (HEC) Program through the NASA Advanced Supercomputing (NAS) Division at Ames Research Center.  We are grateful to Andrei Gruzinov, Hendrik van Eerten, Eliot Quataert and Carles Badenes for helpful comments and discussions.

\bibliographystyle{apj}

\begin{thebibliography}{}
\expandafter\ifx\csname natexlab\endcsname\relax\def\natexlab#1{#1}\fi

\bibitem[{{Allard}(2012)}]{2012APh....39...33A}
{Allard}, D. 2012, Astroparticle Physics, 39, 33

\bibitem[{{Aloy} {et~al.}(2000){Aloy}, {M{\"u}ller}, {Ib{\'a}{\~n}ez},
  {Mart{\'{\i}}}, \& {MacFadyen}}]{2000ApJ...531L.119A}
{Aloy}, M.~A., {M{\"u}ller}, E., {Ib{\'a}{\~n}ez}, J.~M., {Mart{\'{\i}}},
  J.~M., \& {MacFadyen}, A. 2000, \apjl, 531, L119

\bibitem[{{Beresnyak}(2012)}]{2012PhRvL.108c5002B}
{Beresnyak}, A. 2012, Physical Review Letters, 108, 035002

\bibitem[{{Blondin} \& {Ellison}(2001)}]{2001ApJ...560..244B}
{Blondin}, J.~M., \& {Ellison}, D.~C. 2001, \apj, 560, 244

\bibitem[{{Bloom} {et~al.}(2009){Bloom}, {Perley}, {Li}, {Butler}, {Miller},
  {Kocevski}, {Kann}, {Foley}, {Chen}, {Filippenko}, {Starr}, {Macomber},
  {Prochaska}, {Chornock}, {Poznanski}, {Klose}, {Skrutskie}, {Lopez}, {Hall},
  {Glazebrook}, \& {Blake}}]{2009ApJ...691..723B}
{Bloom}, J.~S., {Perley}, D.~A., {Li}, W., {et~al.} 2009, \apj, 691, 723

\bibitem[{{Chevalier} {et~al.}(1992){Chevalier}, {Blondin}, \&
  {Emmering}}]{1992ApJ...392..118C}
{Chevalier}, R.~A., {Blondin}, J.~M., \& {Emmering}, R.~T. 1992, \apj, 392, 118

\bibitem[{{De Colle} {et~al.}(2012){De Colle}, {Ramirez-Ruiz}, {Granot}, \&
  {Lopez-Camara}}]{2012ApJ...751...57D}
{De Colle}, F., {Ramirez-Ruiz}, E., {Granot}, J., \& {Lopez-Camara}, D. 2012,
  \apj, 751, 57

\bibitem[{{Duffell} \& {MacFadyen}(2011)}]{2011ApJS..197...15D}
{Duffell}, P.~C., \& {MacFadyen}, A.~I. 2011, \apjs, 197, 15

\bibitem[{{Duffell} \& {MacFadyen}(2013{\natexlab{a}})}]{2013ApJ...776L...9D}
---. 2013{\natexlab{a}}, \apjl, 776, L9

\bibitem[{{Duffell} \& {MacFadyen}(2013{\natexlab{b}})}]{2013ApJ...775...87D}
---. 2013{\natexlab{b}}, \apj, 775, 87

\bibitem[{{Ferrand} {et~al.}(2010){Ferrand}, {Decourchelle}, {Ballet},
  {Teyssier}, \& {Fraschetti}}]{2010AnA...509L..10F}
{Ferrand}, G., {Decourchelle}, A., {Ballet}, J., {Teyssier}, R., \&
  {Fraschetti}, F. 2010, \aap, 509, L10

\bibitem[{{Fraschetti} {et~al.}(2010){Fraschetti}, {Teyssier}, {Ballet}, \&
  {Decourchelle}}]{2010AnA...515A.104F}
{Fraschetti}, F., {Teyssier}, R., {Ballet}, J., \& {Decourchelle}, A. 2010,
  \aap, 515, A104

\bibitem[{{Goodman} \& {MacFadyen}(2008)}]{2008JFM...604..325G}
{Goodman}, J., \& {MacFadyen}, A. 2008, Journal of Fluid Mechanics, 604, 325

\bibitem[{{Granot}(2007)}]{2007RMxAC..27..140G}
{Granot}, J. 2007, in Revista Mexicana de Astronomia y Astrofisica Conference
  Series, Vol.~27, Revista Mexicana de Astronomia y Astrofisica, vol. 27,
  140--165

\bibitem[{{Granot} {et~al.}(2001){Granot}, {Miller}, {Piran}, {Suen}, \&
  {Hughes}}]{2001grba.conf..312G}
{Granot}, J., {Miller}, M., {Piran}, T., {Suen}, W.~M., \& {Hughes}, P.~A.
  2001, in Gamma-ray Bursts in the Afterglow Era, ed. E.~{Costa},
  F.~{Frontera}, \& J.~{Hjorth}, 312

\bibitem[{{Gruzinov}(2000)}]{2000astro.ph.12364G}
{Gruzinov}, A. 2000, ArXiv Astrophysics e-prints, astro-ph/0012364

\bibitem[{{Haugen} {et~al.}(2003){Haugen}, {Brandenburg}, \&
  {Dobler}}]{2003ApJ...597L.141H}
{Haugen}, N.~E.~L., {Brandenburg}, A., \& {Dobler}, W. 2003, \apjl, 597, L141

\bibitem[{{Helder} {et~al.}(2012){Helder}, {Vink}, {Bykov}, {Ohira}, {Raymond},
  \& {Terrier}}]{2012SSRv..173..369H}
{Helder}, E.~A., {Vink}, J., {Bykov}, A.~M., {et~al.} 2012, \ssr, 173, 369

\bibitem[{{Jun} \& {Norman}(1996)}]{1996ApJ...465..800J}
{Jun}, B.-I., \& {Norman}, M.~L. 1996, \apj, 465, 800

\bibitem[{{Kobayashi} {et~al.}(1999){Kobayashi}, {Piran}, \&
  {Sari}}]{1999ApJ...513..669K}
{Kobayashi}, S., {Piran}, T., \& {Sari}, R. 1999, \apj, 513, 669

\bibitem[{{Komissarov} \& {Barkov}(2009)}]{2009MNRAS.397.1153K}
{Komissarov}, S.~S., \& {Barkov}, M.~V. 2009, \mnras, 397, 1153

\bibitem[{{Kosenko} {et~al.}(2011){Kosenko}, {Blinnikov}, \&
  {Vink}}]{2011AnA...532A.114K}
{Kosenko}, D., {Blinnikov}, S.~I., \& {Vink}, J. 2011, \aap, 532, A114

\bibitem[{{Kumar} \& {Panaitescu}(2000)}]{2000ApJ...541L...9K}
{Kumar}, P., \& {Panaitescu}, A. 2000, \apjl, 541, L9

\bibitem[{{Levinson}(2009)}]{2009ApJ...705L.213L}
{Levinson}, A. 2009, \apjl, 705, L213

\bibitem[{{Levinson}(2010)}]{2010GApFD.104...85L}
---. 2010, Geophysical and Astrophysical Fluid Dynamics, 104, 85

\bibitem[{{L{\'o}pez-C{\'a}mara} {et~al.}(2013){L{\'o}pez-C{\'a}mara},
  {Morsony}, {Begelman}, \& {Lazzati}}]{2013ApJ...767...19L}
{L{\'o}pez-C{\'a}mara}, D., {Morsony}, B.~J., {Begelman}, M.~C., \& {Lazzati},
  D. 2013, \apj, 767, 19

\bibitem[{{Lyutikov} \& {Blandford}(2002)}]{2002bjgr.conf..146L}
{Lyutikov}, M., \& {Blandford}, R. 2002, in Beaming and Jets in Gamma Ray
  Bursts, ed. R.~{Ouyed}, 146

\bibitem[{{MacFadyen} \& {Woosley}(1999)}]{1999ApJ...524..262M}
{MacFadyen}, A.~I., \& {Woosley}, S.~E. 1999, \apj, 524, 262

\bibitem[{{McKinney} {et~al.}(2012){McKinney}, {Tchekhovskoy}, \&
  {Blandford}}]{2012MNRAS.423.3083M}
{McKinney}, J.~C., {Tchekhovskoy}, A., \& {Blandford}, R.~D. 2012, \mnras, 423,
  3083

\bibitem[{{Morsony} {et~al.}(2007){Morsony}, {Lazzati}, \&
  {Begelman}}]{2007ApJ...665..569M}
{Morsony}, B.~J., {Lazzati}, D., \& {Begelman}, M.~C. 2007, \apj, 665, 569

\bibitem[{{Nakamura} \& {Shigeyama}(2006)}]{2006ApJ...645..431N}
{Nakamura}, K., \& {Shigeyama}, T. 2006, \apj, 645, 431

\bibitem[{{Nousek} {et~al.}(2006){Nousek}, {Kouveliotou}, {Grupe}, {Page},
  {Granot}, {Ramirez-Ruiz}, {Patel}, {Burrows}, {Mangano}, {Barthelmy},
  {Beardmore}, {Campana}, {Capalbi}, {Chincarini}, {Cusumano}, {Falcone},
  {Gehrels}, {Giommi}, {Goad}, {Godet}, {Hurkett}, {Kennea}, {Moretti},
  {O'Brien}, {Osborne}, {Romano}, {Tagliaferri}, \&
  {Wells}}]{2006ApJ...642..389N}
{Nousek}, J.~A., {Kouveliotou}, C., {Grupe}, D., {et~al.} 2006, \apj, 642, 389

\bibitem[{{Orlando} {et~al.}(2012){Orlando}, {Bocchino}, {Miceli}, {Petruk}, \&
  {Pumo}}]{2012ApJ...749..156O}
{Orlando}, S., {Bocchino}, F., {Miceli}, M., {Petruk}, O., \& {Pumo}, M.~L.
  2012, \apj, 749, 156

\bibitem[{{Panaitescu} \& {Kumar}(2002)}]{2002ApJ...571..779P}
{Panaitescu}, A., \& {Kumar}, P. 2002, \apj, 571, 779

\bibitem[{{Pedersen} {et~al.}(1998){Pedersen}, {Jaunsen}, {Grav}, {Ostensen},
  {Andersen}, {Wold}, {Kristen}, {Broeils}, {Naeslund}, {Fransson}, {Lacy},
  {Castro-Tirado}, {Gorosabel}, {Rodriguez Espinosa}, {Perez}, {Wolf},
  {Fockenbrock}, {Hjorth}, {Muhli}, {Hakala}, {Piro}, {Feroci}, {Costa},
  {Nicastro}, {Palazzi}, {Frontera}, {Monaldi}, \&
  {Heise}}]{1998ApJ...496..311P}
{Pedersen}, H., {Jaunsen}, A.~O., {Grav}, T., {et~al.} 1998, \apj, 496, 311

\bibitem[{{Peng} {et~al.}(2005){Peng}, {K{\"o}nigl}, \&
  {Granot}}]{2005ApJ...626..966P}
{Peng}, F., {K{\"o}nigl}, A., \& {Granot}, J. 2005, \apj, 626, 966

\bibitem[{{Racusin} {et~al.}(2008){Racusin}, {Karpov}, {Sokolowski}, {Granot},
  {Wu}, {Pal'Shin}, {Covino}, {van der Horst}, {Oates}, {Schady}, {Smith},
  {Cummings}, {Starling}, {Piotrowski}, {Zhang}, {Evans}, {Holland}, {Malek},
  {Page}, {Vetere}, {Margutti}, {Guidorzi}, {Kamble}, {Curran}, {Beardmore},
  {Kouveliotou}, {Mankiewicz}, {Melandri}, {O'Brien}, {Page}, {Piran},
  {Tanvir}, {Wrochna}, {Aptekar}, {Barthelmy}, {Bartolini}, {Beskin}, {Bondar},
  {Bremer}, {Campana}, {Castro-Tirado}, {Cucchiara}, {Cwiok}, {D'Avanzo},
  {D'Elia}, {Della Valle}, {de Ugarte Postigo}, {Dominik}, {Falcone}, {Fiore},
  {Fox}, {Frederiks}, {Fruchter}, {Fugazza}, {Garrett}, {Gehrels},
  {Golenetskii}, {Gomboc}, {Gorosabel}, {Greco}, {Guarnieri}, {Immler},
  {Jelinek}, {Kasprowicz}, {La Parola}, {Levan}, {Mangano}, {Mazets},
  {Molinari}, {Moretti}, {Nawrocki}, {Oleynik}, {Osborne}, {Pagani}, {Pandey},
  {Paragi}, {Perri}, {Piccioni}, {Ramirez-Ruiz}, {Roming}, {Steele}, {Strom},
  {Testa}, {Tosti}, {Ulanov}, {Wiersema}, {Wijers}, {Winters}, {Zarnecki},
  {Zerbi}, {M{\'e}sz{\'a}ros}, {Chincarini}, \&
  {Burrows}}]{2008Natur.455..183R}
{Racusin}, J.~L., {Karpov}, S.~V., {Sokolowski}, M., {et~al.} 2008, \nat, 455,
  183

\bibitem[{{Ramirez-Ruiz} {et~al.}(2002){Ramirez-Ruiz}, {Celotti}, \&
  {Rees}}]{2002MNRAS.337.1349R}
{Ramirez-Ruiz}, E., {Celotti}, A., \& {Rees}, M.~J. 2002, \mnras, 337, 1349

\bibitem[{{Rhoads}(1999)}]{1999ApJ...525..737R}
{Rhoads}, J.~E. 1999, \apj, 525, 737

\bibitem[{{Schekochihin} {et~al.}(2004){Schekochihin}, {Cowley}, {Taylor},
  {Maron}, \& {McWilliams}}]{2004ApJ...612..276S}
{Schekochihin}, A.~A., {Cowley}, S.~C., {Taylor}, S.~F., {Maron}, J.~L., \&
  {McWilliams}, J.~C. 2004, \apj, 612, 276

\bibitem[{{Sironi} \& {Goodman}(2007)}]{2007ApJ...671.1858S}
{Sironi}, L., \& {Goodman}, J. 2007, \apj, 671, 1858

\bibitem[{{Spitkovsky}(2008)}]{2008ApJ...673L..39S}
{Spitkovsky}, A. 2008, \apjl, 673, L39

\bibitem[{{van Eerten}(2014)}]{2014arXiv1402.5162V}
{van Eerten}, H. 2014, ArXiv e-prints, arXiv:1402.5162

\bibitem[{{van Eerten} {et~al.}(2010){van Eerten}, {Zhang}, \&
  {MacFadyen}}]{2010ApJ...722..235V}
{van Eerten}, H., {Zhang}, W., \& {MacFadyen}, A. 2010, \apj, 722, 235

\bibitem[{{van Eerten} \& {MacFadyen}(2012)}]{2012ApJ...747L..30V}
{van Eerten}, H.~J., \& {MacFadyen}, A.~I. 2012, \apjl, 747, L30

\bibitem[{{V{\"o}lk} {et~al.}(2005){V{\"o}lk}, {Berezhko}, \&
  {Ksenofontov}}]{2005AnA...433..229V}
{V{\"o}lk}, H.~J., {Berezhko}, E.~G., \& {Ksenofontov}, L.~T. 2005, \aap, 433,
  229

\bibitem[{{Wang}(2011)}]{2011MNRAS.415...83W}
{Wang}, C.-Y. 2011, \mnras, 415, 83

\bibitem[{{Zhang} {et~al.}(2006){Zhang}, {Fan}, {Dyks}, {Kobayashi},
  {M{\'e}sz{\'a}ros}, {Burrows}, {Nousek}, \& {Gehrels}}]{2006ApJ...642..354Z}
{Zhang}, B., {Fan}, Y.~Z., {Dyks}, J., {et~al.} 2006, \apj, 642, 354

\bibitem[{{Zhang} {et~al.}(2003){Zhang}, {Woosley}, \&
  {MacFadyen}}]{2003ApJ...586..356Z}
{Zhang}, W., {Woosley}, S.~E., \& {MacFadyen}, A.~I. 2003, \apj, 586, 356

\bibitem[{{Zrake} \& {MacFadyen}(2013)}]{2013ApJ...769L..29Z}
{Zrake}, J., \& {MacFadyen}, A.~I. 2013, \apjl, 769, L29

\end{thebibliography}

\end{document}